# OPTIMAL DESIGN OF THREE-PLANETARY-GEAR POWER-SPLIT HYBRID POWERTRAINS

W. ZHUANG[1, 2], X. ZHANG[2*], D. ZHAO[2], H. PENG[2] and L. WANG[1]

[1]School of Mechanical Engineering, Nanjing University of Science & Technology, Nanjing 210094, China
[2]Department of Mechanical Engineering, University of Michigan, Ann Arbor 48105, United States of America

**ABSTRACT**– Many of today's power-split hybrid electric vehicles (HEVs) utilize planetary gears (PGs) to connect the powertrain elements together. Recent power-split HEVs tend to use two PGs and some of them have multiple modes to achieve better fuel economy and driving performance. Looking to the future, hybrid powertrain technologies must be enhanced to design hybrid light trucks. For light trucks, the need for multi-mode and more PGs is stronger, to achieve the required performance. To systematically explore all the possible designs of multi-mode HEVs with three PGs, an efficient searching and optimization methodology is proposed. All possible clutch topology and modes for one existing configuration that uses three PGs were exhaustively searched. The launching performance is first used to screen out designs that fail to satisfy the required launching performance. A near-optimal and computationally efficient energy management strategy was then employed to identify designs that achieve good fuel economy. The proposed design process successfully identify 8 designs that achieve better launching performance and better fuel economy, while using fewer number of clutches than the benchmark and a patented design.

**KEY WORDS**: Hybrid electric vehicle, Configuration optimization, Optimal design, Automated modeling

## NOMENCLATURE

| | |
|---|---|
| $A_0$ | Characteristic matrix of one specific configuration |
| $A^*$ | Characteristic matrix of one specific mode |
| DP | Dynamic programming |
| DoF | Degree of freedom |
| ECVT | Electronic Continuous Variable Transmission |
| FUDS | EPA federal urban driving schedule |
| $f_{fuel}$ | Fuel consumption of each step |
| HEV | Hybrid Electric Vehicle |
| HWFET | EPA Highway Fuel Economy Driving Schedule |
| J | Cost function |
| M | Transform matrix |
| $M_{all}$ | Set of all modes |
| $M_{backward}$ | Set of the engine-on backward driving mode |
| $M_{ECVT}$ | Set of the ECVT mode |
| $Mode_{shift}$ | Cost penalty for the mode shift |
| MG : | Motor/generator |
| $N_{mode}$ | Number of modes |
| $N_{design}$ | Number of designs |
| $N_{clutch}$ | Number of possible clutches |
| $N_{conf}$ | Number of configurations |
| $N_p$ | Number of planetary gears |
| P | Transform matrix |
| PEARS | Power-weighted efficiency analysis for rapid sizing |
| PG | Planetary Gear |
| $P_{EV}^{loss}$ | Power loss of the EV mode |
| $P_{EV}^{in}$ | Total battery output power of the EV mode |
| $P_{e\_1}$ | Engine power flowing through the generator to the battery |
| $P_{e\_2}$ | Engine power flowing through the generator to the motor |
| $P_{e\_3}$ | Engine power directly flowing to the vehicle |
| $P_{batt}$ | Battery power which powers the motor |
| $P_{fuel}$ | Energy rate of the fuel injected |
| STC | Speed and torque cell |
| $SOC_{desired}$ | Desired final state of charge of the battery |
| $SOC_f$ | Actual final state of charge of the battery |
| T | Torque, N |
| US06 | EPA high acceleration driving schedule |
| $\delta_{e\_max}$ | Highest efficiency of the engine |
| $\delta_{MG2\_max}$ | Highest efficiency of the MG2 |
| $\delta$ | Efficiency |

* *Corresponding author*. e-mail: Xiaowuz@umich.edu

| $\delta_{MG1\_max}$ | Highest efficiency of the MG1 |
| $\omega$ | Rotational speed, rad/s |
| $\alpha$ | Weighing factor |
| $\beta$ | Weighing factor |
| $\mu$ | A flag that indicates whether the battery assist is on |

SUBSCRIPTS

| EV | Electric drive mode |
| hybrid | Hybrid driving mode |
| e | Internal combustion engine |
| MG1 | Motor/generator 1 |
| MG2 | Motor/generator 2 |

## 1. INTRODUCTION

Since 1999, more than 3.5 million hybrid electric vehicles have been sold in the US (Jeff et al, 2014). Hybrid sales in 2014 are more than 450,000 and are expected to keep increasing (Jeff, 2014) due to future CAFÉ requirements.

There are three types of HEVs: series, parallel and power-split. The majority of HEVs sold are the power-split type which combines the benefits of series and parallel hybrid powertrains (Jeff et al, 2014). A key feature of power split HEVs using PGs is the Electronic Continuous Variable Transmission (ECVT) which enables efficient engine operations. The two electric machines help to decouple the engine from the road load and speed. If the power devices are sized and controlled well, excellent fuel economy and driving performance can be achieved simultaneously (Wang & Frank, 2014).

Single-PG ECVT is on the extreme end of hardware simplicity. Three of the top selling HEV designs available today, Toyota Prius, Chevrolet Volt and Ford Fusion, all started with single-PG designs. Some other HEV models from Lexus and General Motors use two or even three PGs (Holmes & Schmidt, 2002; Klemen & Schmidt, 2002). When multiple PGs are used, those powertrains can have a compound-split mode which has two mechanical points and theoretically have better fuel economy, especially at higher vehicle speeds. This is in contrast to input and output split modes, both with only one mechanical point (Miller, 2006). The best hybrid powertrain design depends on the vehicle load and driving cycles. Combat vehicles (Liu & Peng, 2010) and urban delivery trucks (Li & Peng, 2010) may require different designs than highway vehicles.

Recently, multi-mode hybrid configurations, realized by adding clutches, were introduced. By switching the clutch states, different operating modes can be achieved. ECVT modes (input-split, output-split and compound-split), parallel modes, series modes, battery electric modes and parallel fixed gear modes could all be achieved in the same powertrain. The existence of multiple operating modes makes it possible to achieve high fuel economy and driving performance than single-mode hybrids. For example, input-split mode can be operated at lower vehicle speeds to achieve better launching performance, while parallel fixed-gear mode may be very efficient in highway driving. Two example multi-mode HEVs available on the market today are Chevrolet Volt and Honda Accord.

Several multi-mode HEV designs were studied recently. Zhang et al. investigated the Toyota Prius architecture and proposed a new design with an additional clutch to achieve considerable fuel economy improvement (Zhang et al, 2012). A hybrid Chevy Tahoe was introduced in (Grewe et al, 2007), this "two-mode hybrid" has four parallel fixed-gear modes and two ECVT modes, higher efficiency through a set of drive cycles was achieved (Kim et al, 2011).

All feasible configurations and designs using a single planetary gear have been exhaustively searched (Zhang et al, 2013). A similar study was conducted using the bond graph technique (Bayrak et al, 2013). The number of configurations with a single PG is small and all possible configurations can be easily explored. For 2PGs or 3PGs hybrids with multiple clutches, the number of valid designs is much larger. An efficient search and optimization process is thus needed and one possible approach is proposed in this paper. Its effectiveness is demonstrated in a case study.

This paper is organized as follows: In Section 2, the GM 2-mode hybrid powertrain is introduced and a near-optimal energy management method is presented and used in its analysis. In Section 3, an automated modeling technology is introduced to generate all possible modes for all designs with the design space. Section 4 describes the methodology for searching and optimizing the designs. Finally, conclusions are presented in Section 5.

## 2. GM 2-MODE HYBRID POWERTRAIN

A "two-mode hybrid" was introduced by GM in (Raghavan et al, 2007). It has four parallel fixed-gear modes, an input-split mode, and a compound-split mode. It was based on the design shown in (Schmidt & Klemen, 2000), can satisfy the high acceleration, hill climbing and towing needs, and had been applied to product hybrid GMC Yukon and Chevrolet Tahoe. Fig. 2 shows the lever diagram of this design, which consists of three PGs, two motor/generators (MGs), and four clutches. Table 1 shows the clutch states for these modes. The parallel fixed-gear modes (with gear ratio between 0.75 and 3) were introduced to improve the fuel economy as well as launching performance.

The benefits of having the parallel fixed-gear modes are investigated below. To make a fair comparison, another design is introduced: the "original 2-ECVT hybrid", which has the same configuration as the GM 2-mode hybrid but has only two clutches, clutch 1 and 3 in Fig. 2, and only two ECVT modes. A near-optimal control algorithm is employed here, called Power-weighted Efficiency Analysis for Rapid Sizing (PEARS) (Zhang et al, 2013). The results generated by PEARS are close to what can be achieved through Dynamic Programming (DP), yet the computation time is much faster. PEARS also has better performance than other energy management strategy, like PMP (Kim et al, 2014), ECMS (Zheng et al, 2014) and so on (Yan et al, 2014). In this section, we will describe the procedure of PEARS and compare the GM 2-mode hybrid and the original 2-ECVT hybrid without parallel fixed gear modes.

### 2.1. Procedure of PEARS+

In HEVs, the engine power can flow through either the electrical path or the mechanical path to drive the vehicle. In the electric path, the battery can help the engine to drive the vehicle or absorb the excess engine power. To minimize the overall losses, a weighted overall efficiency will be analyzed via an algorithm named the Power-weighted Efficiency Analysis for Rapid Sizing (PEARS) (Zhang et al, 2013). In this paper, an improved version will be presented which utilizes DP to find the optimal mode shift schedule. This modified version is called the PEARS+ method. Its process is shown in Fig. 1 and explained below.

#### 2.1.1. PEARS Analysis

a) For a given driving cycle, the speed and road load data are extracted and discretized into vehicle speed-torque cells. Each cell is called a Speed and Torque cells (STC) for subsequent discussion;
b) After formulating the STCs of a given driving cycle, all possible EV modes and hybrid modes of the given HEV powertrain are analyzed. The hybrid/EV modes here refer to the modes with/without the engine running. The efficiency of the EV modes can be calculated from Eq. (1), where $P_{EV}^{loss}$ consists of the battery loss and the electric drivetrain loss, and $P_{EV}^{in}$ is the battery power.

$$\delta_{EV} = 1 - \frac{P_{EV}^{loss}}{P_{EV}^{in}} \quad (1)$$

In the hybrid modes, the power flow can be divided into four parts: the engine power flowing through the generator to the battery $P_{e\_1}$, the engine power flowing through the generator to the motor $P_{e\_2}$, the engine power directly flowing to the vehicle $P_{e\_3}$, and the battery power $P_{batt}$ which powers the motor. The power-weighted efficiency is calculated from:

$$\delta_{Hybrid}(\omega_e, T_e) = \frac{P_{e\_1}\delta_{MG2}\delta_{batt}/(\delta_{e\_max}\delta_{MG2\_max})}{P_{fuel}+\mu P_{batt}} + \frac{P_{e\_2}\delta_{MG1}\delta_{MG2}/(\delta_{e\_max}\delta_{MG1\_max}\delta_{MG2\_max})}{P_{fuel}+\mu P_{batt}} + \frac{P_{e\_3}/\delta_{e\_max}+\mu P_{batt}\delta_{batt}\delta_{MG2}/\delta_{MG2\_max}}{P_{fuel}+\mu P_{batt}} \quad (2)$$

Table 1 Clutch states of the six modes of the GM 2-mode hybrid vehicle

| Mode | Clutch1 | Clutch2 | Clutch3 | Clutch4 |
|---|---|---|---|---|
| Input-Split | X | | | |
| FG1 | X | | | X |
| FG2 | X | X | | |
| Compound-Split | | X | | |
| FG3 | | X | | X |
| FG4 | | X | X | |

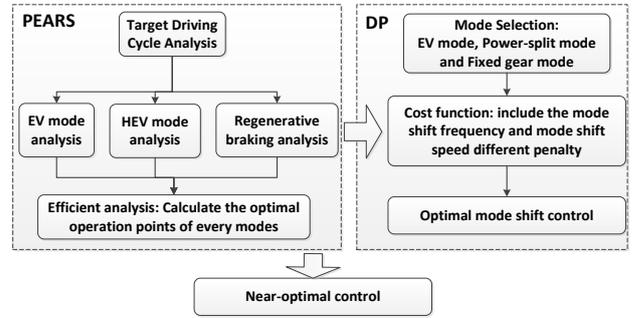

Fig. 1 The procedure of the PEARS+ method

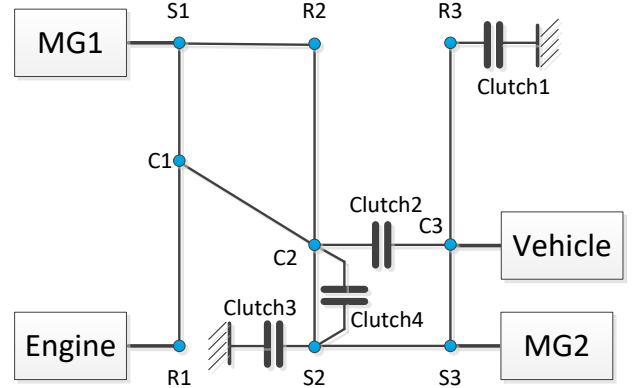

Fig. 2 Lever diagram of the GM 2-mode hybrid vehicle

where $P_{fuel}$ stands for the energy rate of the fuel injected. $\delta_{e\_max}$, $\delta_{MG1\_max}$, $\delta_{MG2\_max}$ are the highest efficiency of the engine, MG1 and MG2, respectively. μ is the flag that indicates whether the battery assist is on: μ = 0 if the battery power is less than 0; μ = 1 if the battery power is greater than or equal to 0.

c) By computing the power-weighted efficiency for each STCs, the highest efficiency operation mode is identified. The torque combinations

($T_{MG1}$ and $T_{MG2}$) of EV modes and the engine states ($\omega_e, T_e$) of Hybrid modes are determined and recorded.

### 2.1.2. DP analysis

The Dynamic Programming technique is used to determine the operating modes under different driving conditions and to calculate the optimal mode trajectory in the driving cycle. The battery state of charge and the operating mode are taken as the state variables. The following cost function is used in the DP problem:

$$J = \min\left[\sum_{k=0}^{N}(f_{fuel}(k) + \alpha Mode_{shift}(k)) + \beta(SOC_{desired} - SOC_f(k))^2\right] \quad (3)$$

where $f_{fuel}(k)$ is fuel consumption at stage k determined by the PEARS analysis. The second term of the cost function is used to avoid frequent mode shifts. Its mathematical form is:

Table 2 the parameters of GM 2-mode hybrid

| Component | Parameter | Value |
|---|---|---|
| Engine | Displacement | V8 6.0 L |
| | Maximum Power | 248 kW @ 5100RPM |
| | Maximum Torque | 498 N•m @ 4100RPM |
| | Inertial | 0.22 kg•m$^2$ |
| MG1, MG2 | Maximum Power | 60Kw |
| | Maximum Torque | 300Nm |
| | Maximum Speed | 9000RPM |
| Ni-MH battery | Maximum Power | 40Kw |
| | Voltage | 300V |
| Planetary gear | Ring/Sun ratio | 2 |
| Final Drive | Gear Ratio | 3.42 |
| Vehicle | Mass | 2680Kg |
| | Tire radius | 0.4m |

$$Mode_{shift}(k) = \alpha_1[\omega_e(k+1) - \omega_e(k)]^2 + \alpha_2[\omega_{MG1}(k+1) - \omega_{MG1}(k)]^2 + \alpha_3[\omega_{MG2}(k+1) - \omega_{MG2}(k)]^2 \quad (4)$$

where $\omega_e(k)$, $\omega_{MG1}(k)$ and $\omega_{MG2}(k)$ are the current rotational speeds of the engine, MG1 and MG2, while $\omega_e(k+1)$, $\omega_{MG1}(k+1)$ and $\omega_{MG2}(k+1)$ are the speeds at the next stage. The third term of the cost function avoids large SOC fluctuation. $\alpha_1$, $\alpha_2$, $\alpha_3$, $\alpha$ and $\beta$ are the weighting factors to be tuned. $\alpha$ is tuned to reduce the mode shift frequency to a reasonable value; $\beta$ is tuned to enforce the battery SOC in the end of the driving cycle back to the original SOC value.

The simplified DP problem plus the pre-computed PEARS table can be used to compute vehicle fuel consumptions and it takes about 1 minute to compute for the 1,372 seconds long Federal Urban Driving Schedule (FUDS).

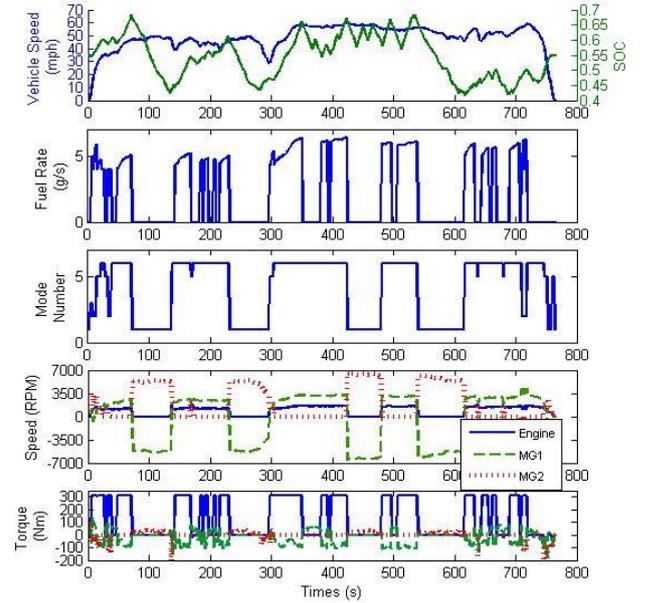

Fig. 3 Simulation results of the GM 2-mode hybrid powertrain in the HWFET cycle

Table 3 Fuel consumption and acceleration comparison

| Driving Cycle | Original 2-mode ECVT mpg | GM 2-mode hybrid mpg | Improvement |
|---|---|---|---|
| FUDS | 29.26 | 29.96 | 2.39% |
| HWFET | 27.04 | 27.70 | 2.44% |
| US06 | 19.38 | 21.32 | 10.01% |
| 0-100kph time (s) | 13.02 | 6.69 | 48.6% |

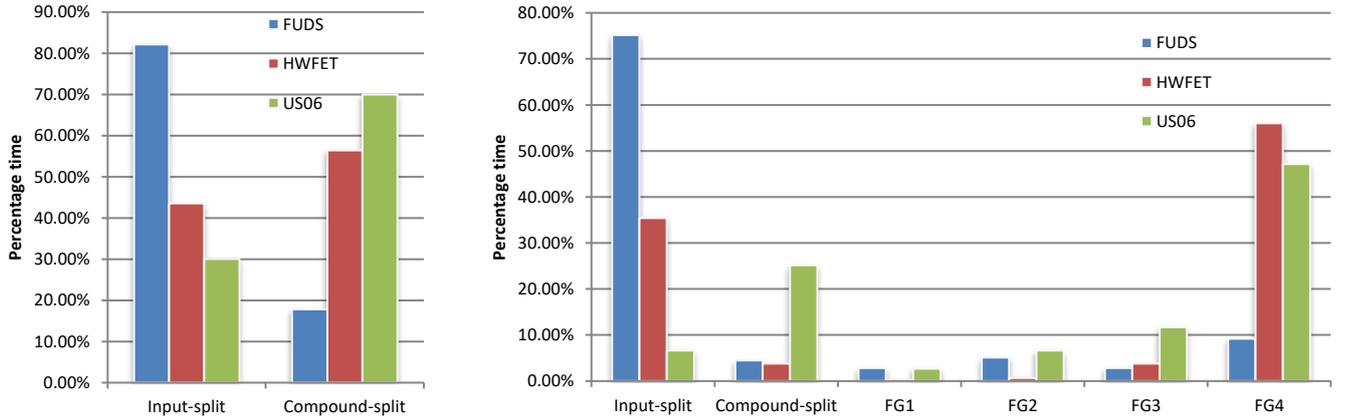

Fig. 4 Mode usage statistics of the two hybrid powertrain

### 2.2. Comparison between the GM 2-mode hybrid and the original 2 ECVT hybrid

The PEARS+ method is used to obtain optimal mode and energy management of the GM 2-mode hybrid and the original 2-mode ECVT hybrid. Table 2 gives the parameters of the GM 2-mode hybrid. Table.3 summarizes the improvement due to the clutches and multi-modes. As shown in Fig. 3, the mode shift does not frequently occur. For the two highway cycles, HWFET and US06, parallel fixed gear mode 3 and parallel fixed gear mode 4 were used 50% of the time based on Fig.5 where shows the usage percentage of each mode under three driving cycles. The fuel consumption in the US06 cycle is reduced significantly by the parallel fixed gear modes. For the city cycle (FUDS), the fuel economy improvement is small.

From the above analysis, the GM 2-mode hybrid adds 2 clutches but only improves fuel economy slightly. We hypothesized that this is because the added hardware complexity (clutches) was not optimally selected. In the following, we aim to find designs with better fuel economy and launching performance than the GM 2-mode hybrid, while using fewer clutches and modes.

In addition, a key issue of the GM 2-mode design is that it cannot use the engine to drive the vehicle backwards. The engine-on backward driving mode is essential for full-size, full-utility SUVs and light trucks, especially when the battery SOC is low. In the following search process, we will require engine-on backward driving as an additional attribute in all winning designs.

## 3. CONFIGURATION AND CLUTCH LOCATION SEARCH

Before explaining the design process, several terms and assumptions are defined. The number of PGs $N_p$ discussed in this paper is 3, and the total number of PG nodes is 9.

### 3.1. Configuration

Configuration refers to the locations of the powertrain components which are the engine, two MGs and the output node to the vehicle drive axle. Each device can connect with any of the 9 nodes. Thus, the total number of configurations is $N_{conf} = P_9^4 = 3{,}024$.

### 3.2. Mode

"Mode" stands for dynamics of the powertrain for a given state of the clutches for a specific configuration. The total number of possible clutches that connect two nodes, or a node with the ground, is:

$$N_{clutch} = C_{3N_p}^2 + 3N_p - 2N_p - 1 \qquad (5)$$

where the first term is the number of clutches that can be added between any two nodes, and the second term stands for the number of total possible grounding clutches. Because locking any two of the three nodes in a PG produce identical dynamics, the third term eliminates the redundant clutches. Finally, the output node should not be grounded. Fig. 5 shows all possible clutch locations for a three PG system. In total, 38 clutches can be added and the $(2N_p + 1)$ redundant clutches are marked in red (assuming the vehicle output is on the 3rd carrier gear).

The degree of freedom of a single planetary gear is two. Therefore, the degree of freedom (DoF) of the 3PGs powertrain without any connection starts from six. Each effective clutch engagement reduces the DoF by one. For Power-split HEVs, there are three controllable components: an engine and two MGs. Thus, the meaningful DoF of the system is 1 (ex. parallel fixed gear mode), 2 (ex. ECVT mode) and 3 (ex. engine speed and one of the MGs speed is free). Therefore, the number of connections by clutches or permanent connections is

between 3 and 5. For any given configuration, the number of total possible modes is

$$N_{mode} = C_{38}^3 + C_{38}^4 + C_{38}^5 = 584193 \quad (6)$$

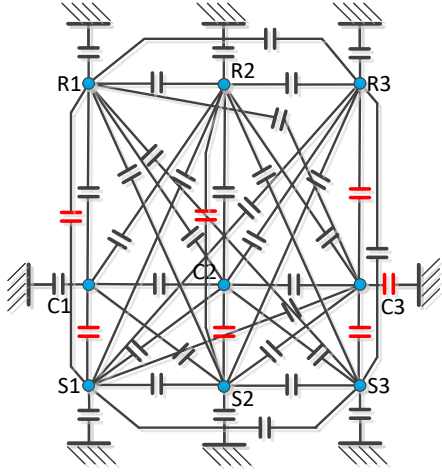

Fig. 5 All possible clutches for a 3PG powertrain

### 3.3. Design

A "Design" represents a configuration with a given number of clutches/permanent connections and their locations. By engaging and disengaging the clutches, multiple modes can be realized in one design. In this paper, we aim to identify designs with better acceleration and fuel economy than the GM 2-mode hybrid with no more than 3 clutches. If we limit ourselves to use 3 permanent connections and 3 clutches, the total number of valid designs is:

$$N_{design} = N_{conf} \times C_{38}^3 \times C_{35}^3 = 1.67 \times 10^{11} \quad (7)$$

Apparently, the number of designs is too tremendous to be evaluated exhaustively. For simplicity, several assumptions are made.
1) We focus on one particular configuration which has the same configuration as the GM 2-mode hybrid as shown in Fig. 6.
2) Out of the three permanent connections, one is between the first and the second PG, and another between the second and the third PG.

With the two assumptions, the number of designs reduces to.

$$N_{design} = (C_3^1 \times C_3^1 \times C_3^1 \times C_3^1 \times C_{38-2}^1) \times C_{35}^3 = 19085220 \quad (8)$$

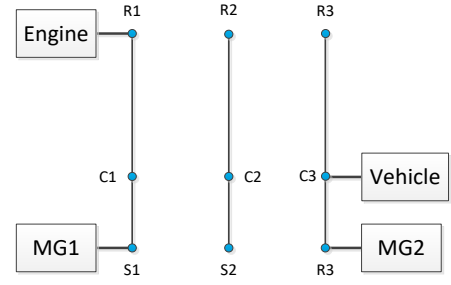

Fig. 6 the configuration that applied in this paper

## 4. DESIGN SEARCH AND OPTIMIZATION

Due to the large size of the design pool, a reliable and fast method is needed to search through all the designs. The process can be divided into two parts: mode analysis and design analysis as shown in Fig. 7. In the mode analysis step, the model of the multi-mode HEV is derived and analyzed automatically. Then, an efficient screening method is used to find the designs that have acceptable launching performance and fuel economy.

### 4.1. Mode Analysis

#### 4.1.1. Automatic Modeling
The flow chart of the automated modeling process is shown in Fig.9. The mode dynamics of any specific mode is governed by the characteristic matrix $A_0$, which is obtained from the configuration and the transformation matrices M and P. More details can be found in (Zhang et al, 2014).

#### 4.1.2. Mode classification

After deriving the dynamics of the mode, the characteristic matrix $A^*$ can be calculated. It governs the relationship between the angular acceleration of the powertrain devices and the torques, as shown in Eq.(9). The details on how to obtain $A^*$ can be found in (Zhang et al, in press).

$$\begin{bmatrix} \dot\omega_{out} \\ \dot\omega_{eng} \\ \dot\omega_{MG1} \\ \dot\omega_{MG2} \end{bmatrix} = A^* \begin{bmatrix} T_{load} \\ T_{eng} \\ T_{MG1} \\ T_{MG2} \end{bmatrix} \quad (9)$$

The characteristic matrix $A^*$ is used to eliminate infeasible and redundant modes afterward. If the vehicle cannot be powered by any devices for a given mode, it is said to be an infeasible mode. In addition, only one of the redundant modes (that share the same $A^*$ matrix ) is kept.

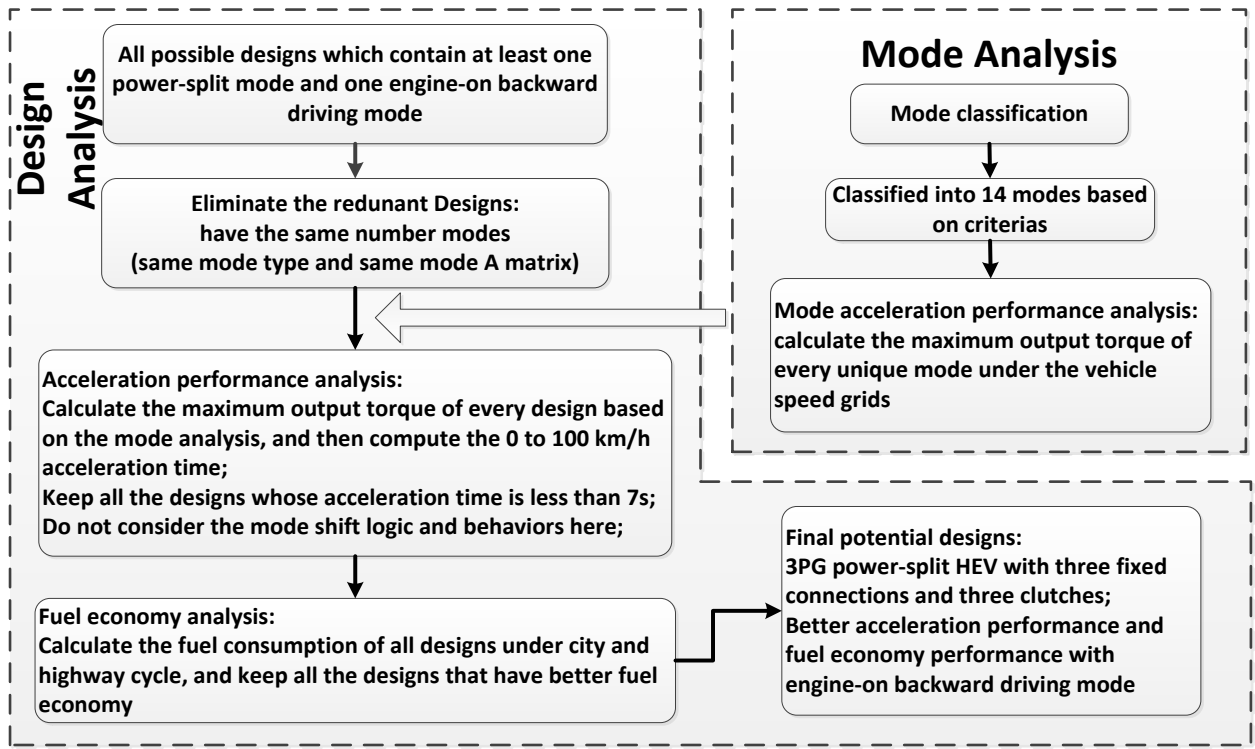

Fig. 7 The searching and optimization process

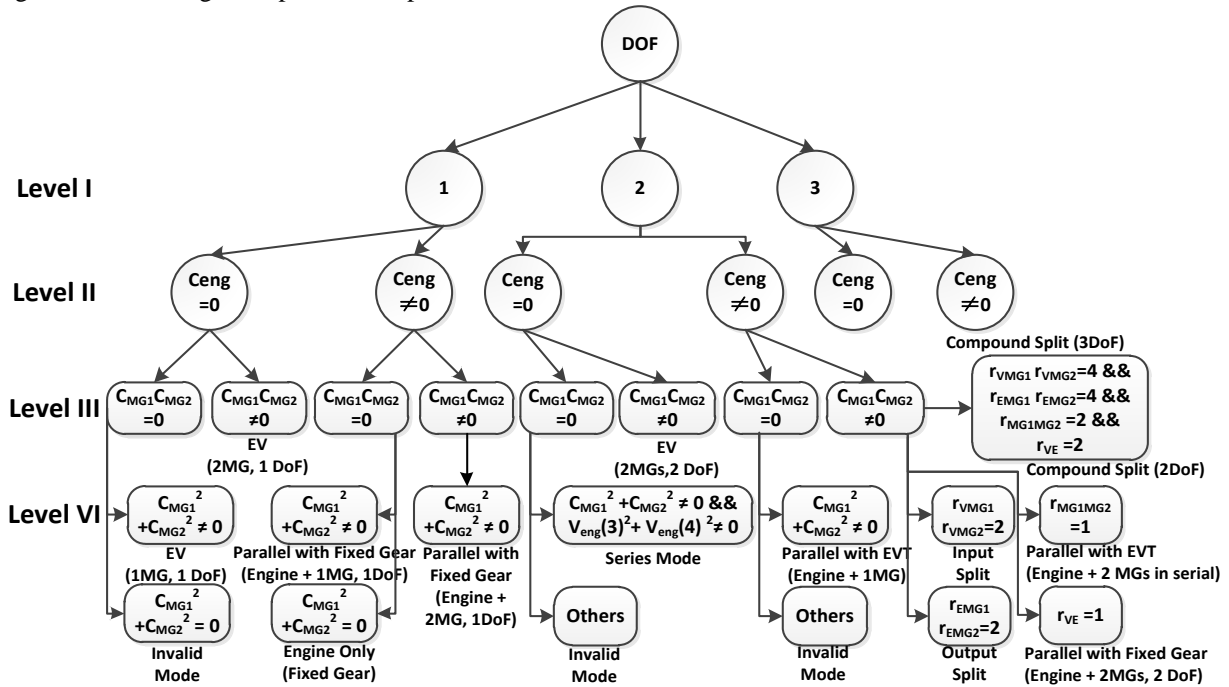

Fig. 8 The binary tree for the mode classification

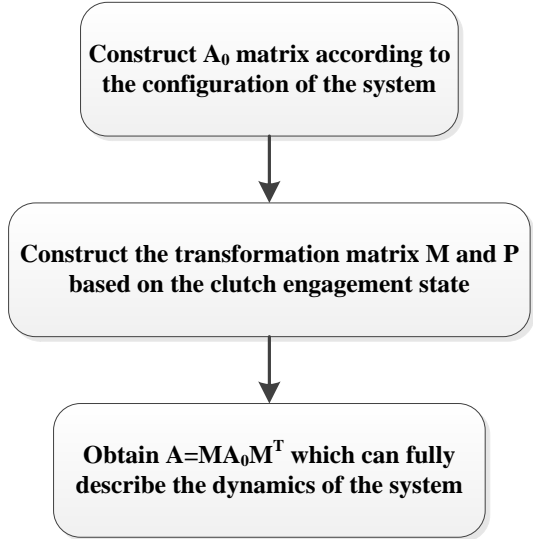

Fig. 9 The automatic modeling process

### 4.1.3. Mode derivation

The rank of the characteristic matrix reflects the DoF of the hybrid powertrain. For all useful modes, the DoF is between one and three. The four rows of the $A^*$ matrix are named as $V_{veh}$, $V_{eng}$, $V_{mg1}$ and $V_{mg2}$, respectively, and the elements of the $V_{veh}$ are named as $C_{veh}$, $C_{eng}$, $C_{mg1}$ and $C_{mg2}$. Additionally, six other matrix are defined: $M_{VE} = [V_{veh}; V_{eng}]$, $M_{vmg1} = [V_{veh}; V_{mg1}]$ $M_{vmg2} = [V_{veh}; V_{mg2}]$, $M_{EMG1} = [V_{eng}; V_{mg1}]$, $M_{emg2} = [V_{eng}; V_{mg2}]$ and $M_{mg1mg2} = [V_{mg1}; V_{mg2}]$. The ranks of these matrices are denoted as $R_{VE}$, $R_{VMG1}$, $R_{VMG2}$, $R_{EMG1}$, $R_{EMG2}$ and $R_{MG1MG2}$.

Based on these matrices, the modes are classified into 14 types. A binary tree is created to classify all the possible modes as shown in Fig. 8. The modes are classified into 14 different types and Table 4 summarizes the criteria for mode classification. The numbers of all the modes are shown in Table. 5. Even though the original mode number is large, the number of unique modes is much smaller. For example, Fig.10 shows an example of several modes sharing the same characteristic matrix. For case (a), selecting any two dash lines on one side of the second PG and any one from the other side results in the same mode.

Define the number of modes in each mode class as $M_1, M_2, \cdots, M_{14}$, then the set of all modes is

$$M_{all} = \{M_1, M_2, \cdots, M_{14}\} \quad (10)$$

The set of engine-on backward driving modes $M_{backward}$ and ECVT modes $M_{ECVT}$ can be described as Eq.(11) and (12). Note that, the series modes $M_1$ are all the engine-on backward driving mode.

$$M_{backward} = \{m_n | m_n(1,2) < 0, m_n \in M_{all}\} \cup M_1 \quad (11)$$
$$M_{ECVT} = \{m_n | m_n(1,2) > 0, m_n \in \{M_3, M_4, M_5\}\} \quad (12)$$

where $m_n$ is any mode in $M_{all}$, and $m_n(1,2)$ is the first row and the second column of the characteristics matrix $A^*$ of the mode $m_n$.

Table 4 Criteria of mode classification

| | Mode Classification | Criteria |
|---|---|---|
| 1 | Series Mode | DoF = 2, $C_{eng} = 0$, $C_{MG1}C_{MG2} = 0$ $V_{eng}(3) V_{eng}(4) = 0$, $C_{MG1}^2 + C_{MG2}^2 \neq 0$ $V_{eng}(3)^2 + V_{eng}(4)^2 \neq 0$ |
| 2 | Compound Split (3 DoF) | DoF = 3 |
| 3 | Compound Split (2 DoF) | DoF = 2, $C_{eng} \neq 0$, $C_{MG1}C_{MG2} \neq 0$, RVE=2, $R_{VMG1}R_{VMG2} = 4$ $R_{EMG1}R_{EMG2} = 4$, $R_{MG1MG2} = 2$ |
| 4 | Input Split | DoF = 2, $C_{eng} \neq 0$, $CMG1CMG2 \neq 0$ $R_{VMG1} R_{VMG2} = 2$ |
| 5 | Output Split | DoF = 2, $C_{eng} \neq 0$, $CMG1CMG2 \neq 0$ $R_{EMG1} R_{EMG2} = 2$ |
| 6 | Parallel with ECVT (Engine + 1MG) | DoF = 2, $C_{eng} \neq 0$, $C_{MG1}C_{MG2} = 0$, $C_{MG1}^2 + C_{MG2}^2 \neq 0$ |
| 7 | Parallel with ECVT (Engine + 2 MGs in serial) | DoF = 2, $C_{eng} \neq 0$, $C_{MG1}C_{MG2} \neq 0$, $R_{MG1MG2} = 1$ |
| 8 | Engine Only (Fixed Gear) | DoF = 1, $C_{eng} \neq 0$, $C_{MG1} C_{MG2} = 0$, $C_{MG1}^2 + C_{MG2}^2 = 0$ |
| 9 | Parallel with Fixed Gear (Engine + 2MGs, 2 DoF) | DoF = 2, $C_{eng} \neq 0$, $R_{VE} = 1$, $C_{MG1}C_{MG2} \neq 0$ |
| 10 | Parallel with Fixed Gear (Engine + 2MGs, 1DoF) | DoF = 1, $C_{eng} \neq 0$, $C_{MG1}C_{MG2} \neq 0$ |
| 11 | Parallel with Fixed Gear (Engine + 1MG, 1DoF) | DoF = 1, $C_{eng} \neq 0$, $C_{MG1} C_{MG2} = 0$, $C_{MG1}^2 + C_{MG2}^2 \neq 0$ |
| 12 | EV (2MGs, 2 DoF) | DoF = 2, $C_{eng} = 0$, $C_{MG1}C_{MG2} \neq 0$ |
| 13 | EV (2MGs, 1 DoF) | DoF = 1, $C_{eng} = 0$, $C_{MG1}C_{MG2} \neq 0$ |
| 14 | EV (1MG, 1 DoF) | DoF = 1, $C_{eng} = 0$, $C_{MG1} C_{MG2} = 0$, $C_{MG1}^2 + C_{MG2}^2 \neq 0$ |

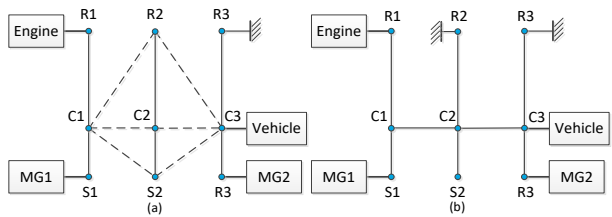

Fig. 10 Example of redundant modes

Table 5 Mode number and unique mode number

| Mode | Original Modes | Unique Modes | Forward Driving Modes |
|---|---|---|---|
| 1 Series Mode | 70,978 | 77 | 77 |
| 2 Compound Split (3 DoF) | 8,830 | 650 | 435 |
| 3 Compound Split (2 DoF) | 2,175 | 207 | 182 |
| 4 Input Split | 12,390 | 188 | 138 |
| 5 Output Split | 13,227 | 186 | 136 |
| 6 Parallel with ECVT (Engine + 1MG) | 2,394 | 94 | 70 |
| 7 Parallel with ECVT (Engine + 2 MGs in serial) | 2,388 | 66 | 54 |
| 8 Engine Only (Fixed Gear) | 10,594 | 31 | 25 |
| 9 Parallel with Fixed Gear (Engine + 2MGs, 2 DoF) | 1,833 | 71 | 65 |
| 10 Parallel with Fixed Gear (Engine + 2MGs, 1DoF) | 240,530 | 972 | 715 |
| 11 Parallel with Fixed Gear (Engine + 1MG, 1DoF) | 68,376 | 504 | 416 |
| 12 EV (2MGs,2 DoF) | 1,279 | 128 | 128 |
| 13 EV (2MGs,1 DoF) | 35,049 | 248 | 248 |
| 14 EV (1MG, 1 DoF,) | 83,527 | 54 | 54 |
| Total | 553,570 | 3,476 | 2,743 |

### 4.2. Design optimization

As shown in Fig.7, the design analysis process is divided into four parts. For designs with 3 permanent connections and 3 clutches, 2DoF and 1DoF modes are achieved with one and two clutch engaging, respectively. 3DoF modes can enable two output shafts for four-wheel drive or skid steering (Pan et al, 2015; Yoshimura, 2013). In this paper, however, we focus on vehicles with only one-output shaft. Therefore, we ignore the 3DoF modes. To sum up, one or two clutches engagement enables at most six modes.

#### 4.2.1 Eliminating inferior designs

Since there are too many designs using 3PG and 3 clutches, two criteria are adopted to screen out inferior designs. The first criterion is that the design should contain at least one power-split mode. The second criterion is that the design must be able to use the engine to drive the vehicle backwards, a feature the GM 2-mode hybrid lacks.

#### 4.2.2 Eliminating redundant designs

Designs that have the same mode number and same mode characteristic matrix are said to be redundant. Fig. 11 shows an example of this situation. The four cases in the figure have different clutch locations, but they share the same dynamics. In other words, they have six same modes. And for this kind design, there are up to 54 designs that are equivalent to each other. After extracting the unique designs, the number of the designs remained is 28,731.

On the basis of this case, it may be meaningfulness to patent every hybrid powertrain design that is manually found. Because others can just change one or two clutch locations to avoid patent infringement but gain the same powertrain performance.

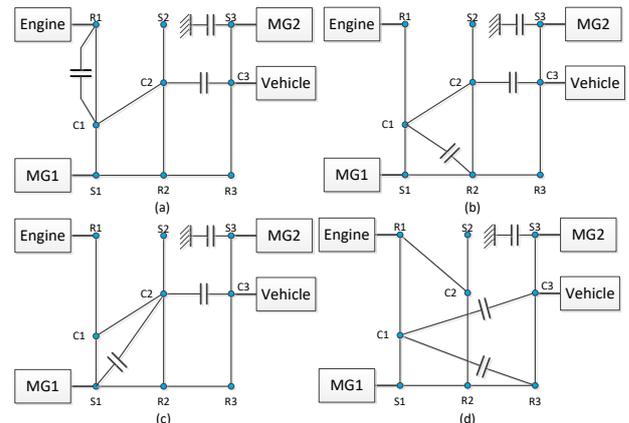

Fig. 11 Example of redundant designs

#### 4.2.3 Acceleration performance evaluation

Good acceleration performance translates to good gradeability and towing capacity, important for SUVs and light trucks. We solve the optimal acceleration problem by computing the maximum output torque of all forward driving modes at all vehicle speeds first. The engine maximum torque is obtained from the engine characteristics curve. The maximum torque of the MGs can be determined using $\omega_{MG1}$ and $\omega_{MG2}$. This computation is performed for all vehicle speeds between 0 and 180 km/h. With these computation done, the maximum acceleration problem of all designs can be solved rapidly.

Fig. 12 shows the distribution of the 0-100km/h acceleration time of all the 28,731 designs that are within the range between 5 seconds and 8 seconds. The y-axis represents the number of designs. Not too many designs accelerate faster than the benchmark—the GM 2-mode hybrid. All designs with acceleration time below 7s are recognized as acceptable and are evaluated for fuel economy. These 261 designs are further divided into two groups: "better acceleration" has a 0-100kph launching in less than 6.69 seconds, and "worse acceleration" has a 0-100kph launching between 6.69 and 7 seconds.

### 4.2.4 Fuel economy evaluation

The PEARS+ methodology is used to calculate the fuel consumption for the city cycle FUDS and highway cycle HWFET. All the powertrain sub-systems other than the "transmission" are the same as those in the GM 2-mode hybrid. Fig. 13 shows the fuel economy results where the black point is the GM 2-mode hybrid, green points indicate the "better acceleration designs" and red points represent the "worse acceleration designs". 29 designs have better fuel economy than GM 2-mode hybrid in both FUDS and HWFET cycles, and 14 of them also have better launching performance. It is important to remember that all of these superior designs use fewer clutches (3 instead of 4) and can drive the vehicle backwards using the engine, which the GM 2-mode hybrid cannot do.

Based on the Environmental Protection Agency's practice, we use 55/45 weights for the FUDS/HWFET cycles to compute the weighted fuel economy (Plug-in Hybrid Electric Vehicles: Learn More About the New Label, 2011). Three designs are selected, named as Designs I, II and III. Design III with 29.38 mpg weighted fuel economy is the best fuel economy design. Fig. 14 and Table 6 show its lever diagram and performance. It has better launching performance as the GM 2-mode but the fuel economy is 2.61% and 0.08% better on the city and highway cycles.

Fig. 15 shows mode usage statistics of Design III. Mode number 1 to 6 stands for input-split mode, output-split backward mode, series mode and parallel fixed gear mode (gear ratio = 1, 1.5, 3), respectively. Note that this design does not contain a pure EV mode, the 'EV' mode statistic shown in Fig. 15 stands for engine-off operation of all six modes. In the city cycle, the 'EV' mode occupies about 85% of time. The input-split mode is used more in the city cycle. On the highway cycle, the fixed-gear modes are used much more frequently, especially for the mode with a torque ratio of 1. It indicates that this low gear ratio mode is most efficient in highway driving. Fig. 16 shows the mode shift behavior in the US06 cycle. The gear will shift to the parallel fixed gear mode (3) in quick acceleration, which has the best acceleration.

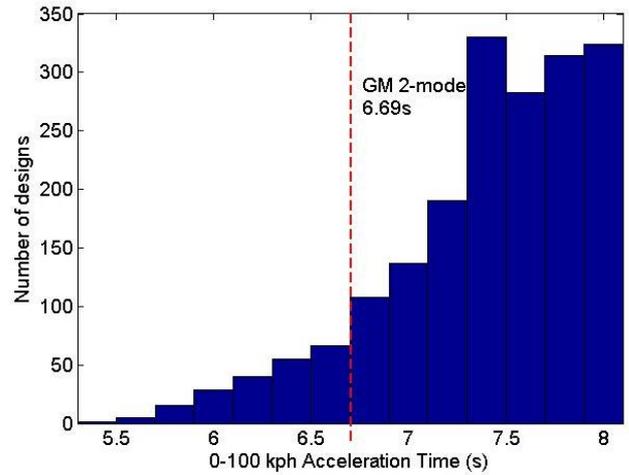

Fig. 12 Acceleration performance of all 28,731 unique designs that fall within the range between 5 seconds and 8 seconds

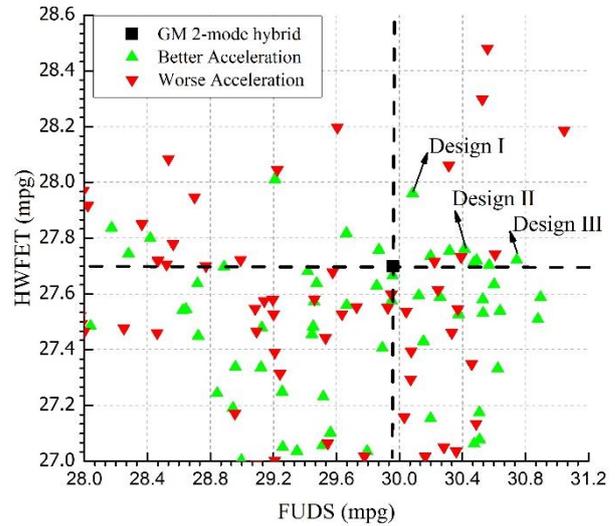

Fig. 13 The fuel economy of the 261 designs that passed the launching performance screening

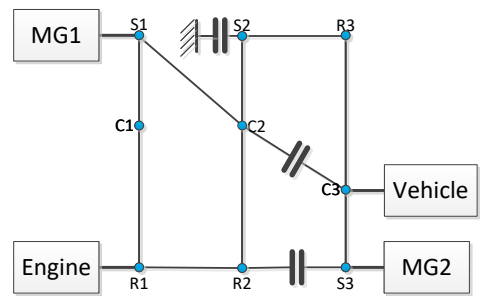

Fig. 14 Identified best 3PG design ("Design III") using three permanent connections and three clutches

Table 6 Mode type and improvement of Design III

| Mode type and number | Improvement |
|---|---|
| 1 input-split mode | FUDS: 2.61% |
| 3 parallel fixed gear forward driving mode (Gear ratio=1, 1.5, 3) | HWFET: 0.08% |
| 1 series mode | 0-100kmh time:6.58s |
| 1 backward output-split mode | |

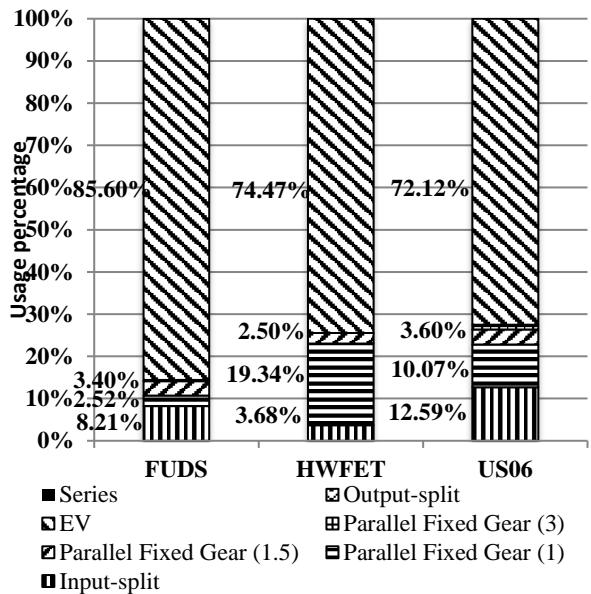

Fig. 15 Usage percentage of the modes for the final winning design shown in Figure 14.

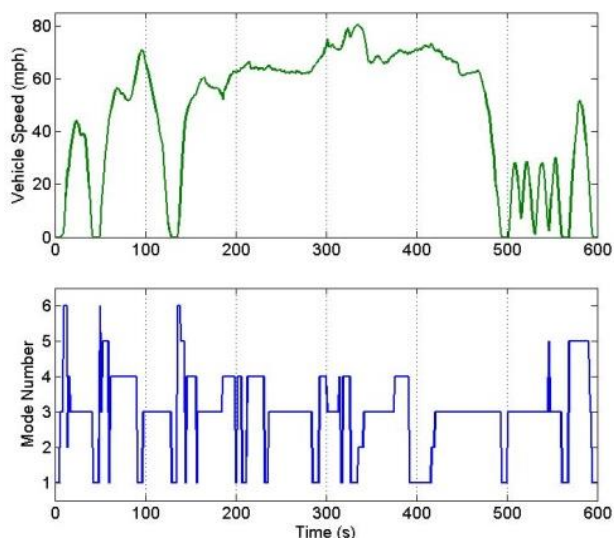

Fig. 16 the mode shift of US06 cycle

## 5. CONCLUSION

In this paper, a design screening process is proposed for exhaustive search of large number of designs in power split hybrid powertrains using three planetary gears and multiple clutches. We first analyze all possible modes, the mode behaviors are then used to study the large number of powertrains designs efficiently. We first use launching performance and engine-on-driving – backwards as criteria to reduce the large design pool to a manageable size. Subsequently, by using a near-optimal control strategy, PEARS+, the fuel economy performance is evaluated. The GM 2-mode hybrid was used as the benchmark. Designs with the same power device connection but different clutch/connection locations are exhaustively searched. At the end we found 14 designs that use fewer clutches, and yet achieve better fuel economy, have better launching performance, and can drive the vehicle backwards with the engine power.

ACKNOWLEDGE

This research is supported by Jiangsu graduate student innovation project, China (KYLX15-0337).